\documentclass[fleqn,10pt]{wlscirep}
\title{Theoretical approach to resonant inelastic x-ray scattering in iron-based superconductors at the energy scale of the superconducting gap}
\usepackage[capitalize]{cleveref}
\usepackage{bm}
\usepackage{bbold}
\usepackage[justification=justified]{caption} 
\newcommand{\meV}{{~\mathrm{meV}}}
\author[1,2,3,*]{Pasquale Marra}
\author[3,4]{Jeroen van den Brink}
\author[3]{Steffen Sykora}
\affil[1]{CNR-SPIN Salerno, I-84084 Fisciano (Salerno), Italy}
\affil[2]{Department of Physics ``E.\,R.\ Caianiello'', University of Salerno, I-84084 Fisciano (Salerno), Italy}
\affil[3]{Institute for Theoretical Solid State Physics, IFW Dresden, D-01069 Dresden, Germany}
\affil[4]{Department of Physics, TU Dresden, D-01062 Dresden, Germany}
\affil[*]{pasquale.marra@spin.cnr.it}


\begin{abstract}
We develop a phenomenological theory to predict the characteristic features of the momentum-dependent scattering amplitude in resonant inelastic x-ray scattering (RIXS) at the energy scale of the superconducting gap in iron-based superconductors.
Taking into account all relevant orbital states as well as their specific content along the Fermi surface we evaluate the charge and spin dynamical structure factors for the compounds LaOFeAs and LiFeAs, based on tight-binding models which are fully consistent with recent angle-resolved photoemission spectroscopy (ARPES) data.
We find a characteristic intensity redistribution between charge and spin dynamical structure factors which discriminates between sign-reversing and sign-preserving quasiparticle excitations.
Consequently, our results show that RIXS spectra can distinguish between $s_\pm$ and $s_{++}$~wave gap functions in the singlet pairing case.
In addition, we find that an analogous intensity redistribution at small momenta can reveal the presence of a chiral $p$-wave triplet pairing.
\end{abstract}
\begin{document}

\flushbottom
\maketitle
\thispagestyle{empty}

\section*{Introduction}

One of the first steps to study the pairing mechanism in unconventional superconductors is the characterization of the superconducting (SC) state with respect to its orbital symmetry and phase.
Whereas in conventional superconductors the superconductivity is based on an effective coupling mediated by phonons in the case of high $T_c$ superconductors one expects electronic correlations to be responsible of the pairing\cite{Leggett2006}, leading to characteristic variations of the energy gap along the Fermi surface\cite{Kirtley1995,Tsuei1997,Damascelli2003}.
Thus, the corresponding order parameter, which describes the momentum dependent coupling strength between the electrons within one Cooper pair, shows a variation in momentum space indicating strong correlations taking place over rather short distances.
Moreover, while the SC gap function of conventional superconductors has the same phase throughout momentum space ($s$-wave symmetry), that of correlation mediated superconductors is expected to exhibit a sign-reversal between Fermi momenta connected by the characteristic wave vector ${\bf Q}_{\rm AF}$ of spin fluctuations\cite{Kuroki2001}.
In $d$-wave superconductors such a sign-change of the SC gap manifests itself in the appearance of gapless quasiparticle excitations that can be detected thermodynamically and by low-energy probes.
However, a sign-reversal of the SC order parameter can also occur, without the presence of nodes, between disconnected hole and electron pockets.
Such a scenario is discussed in the context of an $s_\pm$ symmetry of the SC order parameter in iron-based superconductors\cite{Mazin2008,Kuroki2008,Paglione2010,Stewart2011,Chubukov2012,Hosono2015}.
The relative sign change of the SC gap can be determined, in principle, by phase-sensitive experiments such as Josephson junctions experiments\cite{Tsuei2000}, composite SC loops\cite{Chen2010}, scanning tunneling microscopy (STM)\cite{Hoffman2002,McElroy2003,Hanaguri2007,Kohsaka2008,Hanaguri2009}, and inelastic neutron scattering (INS)\cite{Maier2008,Korshunov2008,Christianson2008,Maier2011,Qiu2008,Inosov2010,Knolle2012}.
Of particular interest in this context is the material LiFeAs\cite{Wang2008,Tapp2008}.
Contrarily to the expected scenario in fact, angle-resolved photoemission spectroscopy (ARPES) have proven the absence of Fermi surface nesting in this compound\cite{Borisenko2010,Borisenko2012}.
Furthermore, the presence of a sign-reversal of the order parameter and the nature of the SC pairing in LiFeAs is still debated.
In fact, theoretical works have given contradictory results, suggesting the presence of either $s$-wave singlet\cite{Platt2011,Wang2013,Ahn2014} (with or without sign-reversal\cite{Kontani2010}) or $p$-wave triplet pairing\cite{Brydon2011}.
On the experimental side, INS experiments\cite{Taylor2011} seem consistent with the presence of spin-singlet pairing with $s$-wave symmetry, whereas STM experiments of the quasiparticle interference\cite{Hanke2012} indicate a $p$-wave spin-triplet state or a singlet pairing mechanism with a more complex order parameter ($s+\imath d$~wave).

In general, the correct interpretation of phase-sensitive experiments requires detailed information about the low-energy properties of the material like, e.g., the normal state band structure, correlations, and the type of impurity scattering.
Moreover, in iron-based superconductors it is well known that the orbital physics and, in particular, the variation of orbital content along the Fermi surface, play a very important role.
In most cases it is the incomplete knowledge about these properties which prevents the characterization of the SC state directly from measured spectra.
One possible way to overcome these problems is by breaking the time-reversal symmetry of the system via an external magnetic field, which is known to enhance the quasiparticle excitations with preserved sign of SC order parameter\cite{Tinkham}.
Recently, this method has been applied to the iron-based superconductor Fe(Se,Te)\cite{Hanaguri2010} and results of this study have been considered as evidence for $s_\pm$ pairing in this material.
However, this interpretation raised a number of comments\cite{CommentHanaguri,ReplyCommentHanaguri}, since, e.g., the field effect in iron-based superconductors is expected to differ significantly from the one observed in cuprates.
Moreover, in this material, the Zeeman splitting is a large fraction of the SC gap size, generating new components to the field induced scattering\cite{Sykora2011}.

Recently, we have shown that RIXS is sensitive to phase and orbital symmetry of SC order parameter, and therefore can be used as an additional experimental method to probe the nature of the SC pairing\cite{Marra2013}.
In the past decade, this spectroscopic technique has been established as an experimental probe of elementary spin\cite{Braicovich2009}, orbital\cite{Ulrich2009}, and lattice excitations\cite{Yavas2010}.
In particular, the direct RIXS process\cite{Ament2011} allows spin flip excitations if the spin of conduction electrons has components parallel to the $x^2-y^2$ orbital\cite{Ament2009}.
This fact offers the possibility to probe the momentum-dependent charge and spin dynamical structure factor (DSF) $\chi^{c,s}({\bf q},\imath\omega)$ of the bulk system\cite{Marra2013}.
We note that the form factors corresponding respectively to charge and spin DSF can be adjusted by the specific geometry of the experiment and the polarization of the incoming photon beam.
This is in contrast with, e.g., INS, where only the spin DSF can be accessed\cite{Maier2008,Korshunov2008,Christianson2008,Maier2011,Qiu2008,Inosov2010,Knolle2012}.
The possibility to probe separately the charge and the spin DSF is therefore a distinctive property of RIXS spectroscopy.
The phase-sensitivity of RIXS, the possibility to probe separately the charge and the spin DSF, and the well known property of spin DSF to enhance sign-reversing scattering (leading to, e.g., the characteristic 41~meV resonant mode in cuprates), would place RIXS experiment on the same footing as INS experiments and STM quasiparticle interference in magnetic field --- with the additional advantage that the interpretation of RIXS spectra does not rely on the modeling of impurity scattering.

The Fe $L_3$-edge RIXS response of iron-bases superconductors has been already studied theoretically\cite{Kaneshita2011} at energies higher than the SC gap, suggesting that RIXS can in principle probe magnon excitations in these systems.
This theoretical work stimulated a subsequent experimental study\cite{Zhou2013}, which not only confirmed the presence of the magnon excitations in an energy range up to an energy of 200~meV in magnetically ordered iron-based superconductors\cite{Harriger2011}, but also revealed the persistence of magnon-like modes in this energy range as the material is doped and becomes SC\@.
It also motivates the question whether and how RIXS can pick up the signatures of the SC gap in these materials.
Probing the SC order parameter requires an energy resolution roughly comparable with twice the magnitude of the gap, i.e., in the order of 10~meV in the iron-based superconductor LiFeAs, which is below the energy resolution of present RIXS facilities\cite{Hancock2010,Zhou2013} at the Fe $L_3$ edge, but in the targeted range of, e.g., the Soft Inelastic X-ray Scattering (SIX) beamline at NSLS-II which is presently under construction.
Nevertheless, the energy resolution of RIXS experiments has improved in the last decade at a very fast pace\cite{Ament2011}, such that one can realistically expect to reach the energy scale of the SC gap in the near future.
On top of that, we notice that RIXS spectra of iron-based superconductors, as well as any other metallic compound, might be influenced by fluorescence.
However, it has been shown that, in order to reliably uncover electronic excitations, the fluorescence background can be subtracted from the spectra\cite{Zhou2013,Hancock2010,Yang2009}.
Moreover, since any spectral feature related to the SC order parameter is obviously absent in the normal state, one can in principle obtain the spectra of SC quasiparticle excitations as the difference between the total inelastic scattering in the normal and in the SC state.
Such differential spectra can be obtained by probing the inelastic response slightly above and sightly below the $T_c$ of the material.
This would have the additional advantage of canceling out not only the fluorescence background, but also any other contribution from excitations which are not related, and thus not affected, by the onset of the SC state.

Motivated by the above considerations, this paper develops a phenomenological approach to the momentum and energy dependent RIXS spectrum in iron-based superconductors at an energy scale of the SC gap, taking into account the relevant orbitals of the iron ions and their different weights along the hole and electron pockets.
In particular, we calculate the charge and spin DSF of the iron-based pnictides LaOFeAs and LiFeAs considering different symmetries for the SC pairing, in the framework of tight-binding models which are able to capture the underlying orbital physics of the compounds and are consistent with ARPES data.
Our study shows that RIXS, being sensitive to the phase of the SC order parameter\cite{Marra2013}, can discriminate between different symmetries of the SC pairing, in particular between singlet $s_\pm$~wave, $s_{++}$~wave, and triplet $p$-wave symmetry.
In fact, RIXS can detect the presence of sign-reversing excitations, which correspond to a sign change of the SC order parameter between hole and electron pocket ($s_\pm$~wave).
These excitations can be disclosed by the presence of a strong intensity peak in the spin DSF at the characteristic wave vector ${\bf Q}_{\rm AF}$ of spin fluctuations.
Moreover, based on these considerations, we present how RIXS could discriminate between singlet and triplet pairing.


\section*{RIXS cross section in iron-based superconductors}

In a \emph{direct} RIXS process at a transition-metal ion $L_{2,3}$ edge, the incoming photon excites a core shell 2$p$ electron into the 3$d$ shell, which consequently decays into an outgoing photon and a charge, spin, or orbital excitation in the electronic system\cite{Ament2011}.
In the case of iron-based superconductors considered in this work, one has to take into account more than one relevant orbital states of the 3$d$ shell.
Note that this multi-orbital structure leads to the characteristic disconnected Fermi surface branches dominating the low-energy properties in these materials.
Using the fast collision approximation\cite{Ament2011,Ament2009,Haverkort2010} the RIXS cross section can be decomposed into a combination of charge and spin DSF of $3d$ electrons as\cite{Marra2013,Kaneshita2011,Haverkort2010,Marra2012}
\begin{equation}
\label{eq:CrossSection}
	I({\bf e},{\bf q},\omega) = \sum\limits_{\alpha\beta}
	|W^c_{\alpha\beta}({\bf e})|^2 \chi^c_{\alpha\beta}({\bf q},\imath\omega) +
	|W^s_{\alpha\beta}({\bf e})|^2 \chi^s_{\alpha\beta}({\bf q},\imath\omega)
	,
\end{equation}
where the charge and spin DSF corresponding to the orbitals $\alpha,\beta$ are defined as
\begin{align}
	\chi^c_{\alpha\beta}({\bf q},\imath\omega) &= \sum\limits_f|\langle f|
	\rho_{\alpha\beta}({\bf q})|i\rangle|^2\delta(\hbar\omega+E_i-E_f),
\nonumber
\\
	\chi^s_{\alpha\beta}({\bf q},\imath\omega) &= \sum\limits_f|\langle f|
	S^z_{\alpha\beta}({\bf q})
	|i\rangle|^2\delta(\hbar\omega+E_i-E_f),
\label{eq:ChargeSpinDSF}
\end{align}
being $|i\rangle$ and $|f\rangle$ the initial and final states of the RIXS process with energy $E_i$ and $E_f$, and with $\hbar\omega$ and ${\bf q}$ the transferred photon energy and momentum.
Note that the spin DSF is assumed to have the same momentum and energy dependence for any direction of the spin\cite{Andersen2005}.
Here, the density and the spin of $3d$ electrons $\rho_{\alpha\beta}({\bf q})=\sum_{{\bf k}\tau} d^{\dagger}_{\alpha\tau{{\bf k}+{\bf q}}} d^{}_{\beta\tau{\bf k}}$ and $S^z_{\alpha\beta}({\bf q})=\sum_{{\bf k}\tau\tau'} d^{\dagger}_{\alpha\tau{{\bf k}+{\bf q}}} \sigma^z_{\tau\tau'} d^{}_{\beta\tau'{\bf k}}$ are defined in terms of the orbital operator $d^{\dag}_{\alpha\tau{\bf k}}$ ($d^{}_{\alpha\tau{\bf k}}$), which creates (annihilates) an electron in the orbital $\alpha$ with spin $\tau$ and momentum $\bf k$.
The RIXS form factors $W^c_{\alpha\beta}({\bf e})$ and $W^s_{\alpha\beta}({\bf e})$ in \cref{eq:CrossSection} depend on the transition-metal ion, the orbital symmetry of the system, the specific geometry of the experiment, and on the polarization ${\bf e}$ of the incoming and outgoing x-ray beams\cite{Ament2009,Haverkort2010,Marra2012}.
Thus, these parameters can be adjusted in the RIXS experiment, and therefore, under construction of particular experimental setups, the cross section will be solely determined either by the charge or by the spin DSF\@.
As it has been shown in Ref.~\citen{Marra2013}, this property can be used to reveal the character of the pairing mechanism in unconventional superconductors.
Motivated by this idea, we study in this paper the charge and spin DSF for iron-based superconductors using accepted band structure models, and comparing different pairing mechanisms and order parameter symmetries.

In order to reproduce correctly the characteristic disconnected Fermi surface of iron-based superconductors, a minimal model for these systems must include more than one $3d$ orbital state on the Fermi surface.
Therefore, a phenomenological description of the unconventional SC state in iron-based superconductors can be achieved by a generalized multi-band mean-field Hamiltonian in the form\cite{Tinkham}
\begin{equation}
\label{eq:Hamiltonian}
	{\cal H} = \sum\limits_{i\tau{\bf k}} \varepsilon_{i{\bf k}} \,
	c^\dag_{i\tau{\bf k}}c^{}_{i\tau{\bf k}}
	-
	\frac12 \sum\limits_{i\tau{\bf k}} \xi_\tau
	\left( \Delta_{\bf k} c^\dag_{i\tau{\bf k}}c^\dag_{i-\tau-{\bf k}} + \Delta_{\bf k}^*
	c^{}_{i-\tau-{\bf k}} c^{}_{i\tau{\bf k}} \right),
\end{equation}
where the operator $c^\dag_{i\tau{\bf k}}$ ($c^{}_{i\tau{\bf k}}$) creates (annihilates) an electron with spin $\tau$ in the energy band $i$, which is described by the bare electron dispersion $\varepsilon_{i{\bf k}}$, and with $\Delta_{\bf k}$ the momentum-dependent order parameter.
The second term in \cref{eq:Hamiltonian} is responsible for the SC state, with the pairing character determined by $\xi_\tau$.
The case of $\xi_\tau=\pm1$ for up and down spin describes the spin-singlet pairing, whereas the case $\xi_\tau=1$ for both spin directions leads to a special type of spin-triplet state.
In general, the triplet pairing term is given by $-\frac12\Delta_{{\bf k}\tau\tau'} c^\dag_{i\tau{\bf k}}c^\dag_{i\tau'-{\bf k}} + \mbox{h.c.}$, with a multi-component order parameter of the form $\Delta_{{\bf k}\tau\tau'}=\imath\left[{\bf d}({\bf k})\cdot{\boldsymbol \sigma}\right]\sigma_y$.
However, in this paper we consider only the simplest case $d_x({\bf k})=d_y({\bf k})=0$ and $d_z({\bf k})=\Delta_{\bf k}$, and therefore the gap function simplifies to $\Delta_{{\bf k}\uparrow\uparrow}=\Delta_{{\bf k}\downarrow\downarrow}=0$ and $\Delta_{{\bf k}\uparrow\downarrow}=\Delta_{{\bf k}\downarrow\uparrow}=\Delta_{{\bf k}}$.

To investigate the RIXS cross section given by \cref{eq:CrossSection} we calculate the DSF on the basis of the model Hamiltonian~(\ref{eq:Hamiltonian}) separately for each of the relevant orbitals.
Using the unitary transformation between orbital and energy band representation defined as
\begin{equation}
\label{eq:Transformation}
	c_{i\tau{\bf k}}=\sum\limits_\alpha \lambda_{i\alpha,{\bf k}} \,
	d_{\alpha\tau{\bf k}}
	,
\end{equation}
we rewrite the density and spin operators $\rho_{\alpha\beta}({\bf q})$ and $S^z_{\alpha\beta}({\bf q})$ in \cref{eq:ChargeSpinDSF} in terms of the operators $c^{}_{i\tau{\bf k}}$ and $c_{i\tau{\bf k}}^\dag$ in the band representation.
Note that, in general, the Hamiltonian is not diagonal with respect to the orbital states because the different orbitals can hybridize with each other.
The transformation matrix elements $\lambda_{i\alpha,{\bf k}}$, which describe the orbital content of conduction bands, are obtained diagonalizing the low-energy tight-binding Hamiltonian of the system.
For this purpose, we consider in this paper the tight-binding model in Ref.~\citen{Raghu2008} for the compound LaOFeAs, and a slightly modified tight-binding model based on Ref.~\citen{Brydon2011} for the compound LiFeAs.

Having expressed the density and spin operators in the DSF in terms of the one-particle operators in the band representation, the next step is to diagonalize the Hamiltonian~(\ref{eq:Hamiltonian}) by the Bogoliubov transformation $c_{i\uparrow{\bf k}} = u_{i{\bf k}}^* \gamma_{i\uparrow{\bf k}} - v_{i{\bf k}} \gamma_{i\downarrow -{\bf k}}^\dag$ and $c_{i\downarrow{\bf k}} = u_{i{\bf k}}^*\gamma_{i\downarrow{\bf k}} + v_{i{\bf k}} \gamma_{i\uparrow -{\bf k}}^\dag$, with $|u_{i{\bf k}}|^2=\frac12\left(1+\varepsilon_{i{\bf k}} / E_{i{\bf k}}\right)$, $|v_{i{\bf k}}|^2=\frac12\left(1-\varepsilon_{i{\bf k}} / E_{i{\bf k}}\right)$, and $u_{i{\bf k}}^* v_{i{\bf k}} =\frac12 \Delta_{\bf k} / E_{i{\bf k}}$ for each of the different bands.
This allows us to determine the ground state $|{\rm BCS}\rangle$ and the excitations of the system, in terms of the quasiparticle operators $\gamma_{i\tau{\bf k}}$ and of the quasiparticle dispersion $E_{i{\bf k}}=\sqrt{\varepsilon_{i{\bf k}}^2+\vert\Delta_{\bf k}\vert^2}$.
In a centrosymmetric superconductor at zero temperature, the excited states contributing to DSF have the form $\gamma^\dag_{j\tau{\bf k}+{\bf q}}\gamma^\dag_{i-\tau-{\bf k}}|{\rm BCS}\rangle$ with energy $E_{i{\bf k}}+E_{j{\bf k}+{\bf q}}$.
It follows that the charge and spin DSF $\hat{\chi}^{c,s}({\bf q},\imath\omega)$ of quasiparticle excitations is described by a matrix of intra-orbital ($\alpha=\beta$) and inter-orbital ($\alpha\neq\beta$) components given by
\begin{equation}
\label{eq:StructureFactor}
	\chi^{c,s}_{\alpha\beta}({\bf q},\imath\omega) =
	\sum\limits_{i j {\bf k}}
	\delta\left(\hbar\omega-E_{i{\bf k}}-E_{j{{\bf k}+{\bf q}}}\right)
	|\lambda_{i\alpha,{\bf k}}
	\lambda_{j\beta,{\bf k+q}}|^2
	\left[
	1 \pm
	\frac{{\rm Re}(\Delta_{\bf k}^{}\Delta_{{\bf k}+{\bf q}}^*)
	\mp\varepsilon_{i{\bf k}}\varepsilon_{j{{\bf k}+{\bf q}}}}
	{E_{i{\bf k}} E_{j{{\bf k}+{\bf q}}}}
	\right]
	,
\end{equation}
where $\alpha,\beta$ span the relevant orbitals of the system and the $\pm$ sign distinguishes between charge and spin DSF\cite{Kee1998,Kee1999,Voo2000}.
This result shows that the momentum-dependent DSF of low-energy quasiparticle excitations is strongly affected by the orbital content of bare electrons and the structure of the SC order parameter.

The character of the SC pairing, which is described by the gap function $\Delta_{\bf k}$, arises at energies close to the Fermi level $\hbar\omega \approx \varepsilon_F$.
There, the main contributions to the DSF correspond to excitations close to the Fermi surface, i.e., those which fulfill the condition $\varepsilon_{i{\bf k}}\varepsilon_{j{\bf k + q}} \ll |\Delta_{\bf k}\Delta_{{\bf k} + {\bf q}}|$.
Assuming a phase dependent order parameter in the form $\Delta_{\bf k} =|\Delta_{\bf k}|e^{\imath\phi_{\bf k}}$, the DSF in \cref{eq:StructureFactor} for low-energy excitations becomes approximately
\begin{equation}
\label{eq:CoherenceFactors}
	\chi^{c,s}_{\alpha\beta}({\bf q},\imath\omega)
	\approx
	\sum\limits_{i j {\bf k}}
	\delta(\varepsilon_{i{\bf k}}\varepsilon_{j{\bf k+q}})
	\delta(\hbar\omega-|\Delta_{{\bf k}+{\bf q}}| - |\Delta_{\bf k}|)
	|\lambda_{i\alpha,{\bf k}}
	\lambda_{j\beta,{\bf k+q}}|^2
	\left[
	1 \pm \cos(\phi_{\bf k} - \phi_{{\bf k}+{\bf q}}) 
	\right].
\end{equation}
Hence the DSF is influenced significantly by the order parameter phase $\phi_{\bf k}$ on the Fermi surface.
In particular, the charge DSF is suppressed for sign-reversing ($\phi_{\bf k}-\phi_{\bf k+q}=\pi $), whereas the spin DSF is suppressed for sign-preserving excitations ($\phi_{\bf k}-\phi_{\bf k+q}=0$).

Interactions between the conduction electrons are taken into account within the RPA\@.
The matrix function of the DSF $\hat{\chi}^{c,s}_{\rm RPA}({\bf q},\imath\omega)$ with interactions is related to the matrix function of the bare DSF $\hat{\chi}^{c,s}({\bf q},\imath\omega)$ from \cref{eq:StructureFactor} via the equation
\begin{equation}
\label{eq:RPA}
\hat{\chi}^{c,s}_{\rm RPA}({\bf q},\imath\omega) = \hat{\chi}^{c,s}({\bf q},\imath\omega) \left[\mathbb{1} - \hat{\Gamma}\hat{\chi}^{c,s}({\bf q},\imath\omega) \right]^{-1},
\end{equation}
where $\mathbb{1}$ is the identity matrix and $\hat{\Gamma}=U\mathbb{1}$ the interaction matrix.
Following Ref.~\citen{Raghu2008}, we have chosen the intra-orbital interaction as $U=W/4$, where $W$ is the bandwidth of the relevant bands, and neglected the inter-orbital interaction ($J=0$) in the actual calculations of the DSF of LaOFeAs and LiFeAs.

Since the RIXS form factors $W^c_{\alpha\beta}({\bf e})$ and $W^s_{\alpha\beta}({\bf e})$ in \cref{eq:CrossSection} can be tuned by properly choosing the experimental setup, RIXS can probe both charge and spin DSF\cite{Marra2013}, which is a unique feature among other spectroscopies.
A comparison between the charge and the spin DSF of quasiparticle excitations allows one to disclose the momentum dependence of the magnitude and of the phase of the SC order parameter and, therefore, the underlying symmetry of the pairing mechanism.
In the next Section we will show the predicted RIXS spectra for the LaOFeAs and LiFeAs iron-based superconductors obtained numerically using the theoretical framework described above.

\section*{Numerical Results}
\subsection*{Minimal orbital model}

\begin{figure}[t]
\centering\includegraphics[scale=1.55]{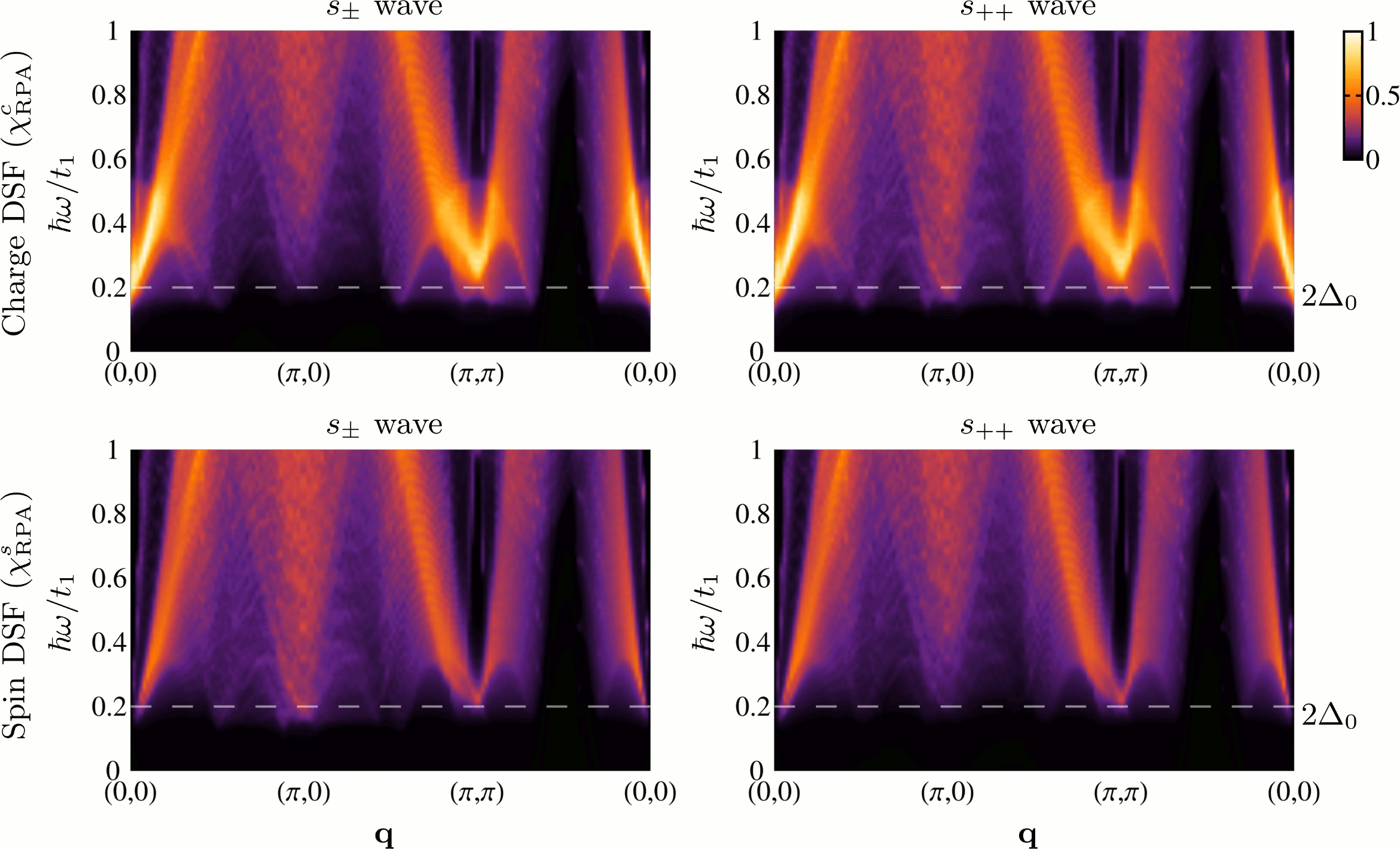}
\caption{
RIXS intensities as a function of energy loss $\hbar\omega$ and transferred momentum $\bf q$ along the high symmetry path in the Brillouin zone, for the charge and spin DSF in LaOFeAs, assuming respectively an $s_\pm$~wave and an $s_{++}$~wave order parameter, calculated via \cref{eq:RPA}, and assuming the bare electron dispersion of the tight-binding model in Ref.~\citen{Raghu2008}.
The resonance peak appears at relatively small energy values close to $(0,0)$ and is dispersive.
Spectral intensities at $(0,0)$ and $(\pi,\pi)$ are suppressed in the spin DSF for both order parameter choices, while spectral intensities at ${\bf Q}_{\rm AF}=(\pi,0)$ are suppressed in the charge (spin) spectra in the $s_\pm$ ($s_{++}$) wave state.
Intensities are in arbitrary units.
}
\label{fig:raghu-confront}
\end{figure}

We start by considering the effective two-band tight-binding model proposed in Ref.~\citen{Raghu2008}, which is regarded as a minimal model for conduction electrons in iron-based superconductors.
This model takes into account the effective hoppings between the two orbitals $xz$ and $yz$ of the iron ions within a single Fe-ion unit cell, and correctly reproduces the band structure of the compound LaOFeAs, which consists of disconnected hole-like Fermi surface branches around the points $(0,0)$ and $(\pi,\pi)$ and separate electron pockets of similar size around the point $(\pi,0)$ in the Brillouin zone.
We assume three different symmetries for the SC gap, i.e., $s_\pm$~wave\cite{Mazin2008,Kuroki2008}, $s_{++}$~wave\cite{Kontani2010}, and a spin-triplet $p_z$~wave\cite{Brydon2011}, where the momentum dependence is modeled respectively by $\Delta^{s_\pm}_{\bf k}=\Delta_0 \cos{k_x}\cos{k_y}$, $\Delta^{s_{++}}_{\bf k}=\vert\Delta^{s_\pm}_{\bf k}\vert$, and $\Delta^{p_z}_{\bf k}=\Delta_0\left(\sin{k_x}-\imath \sin{k_y}\right)$, with $\Delta_0=0.1t_1$, where $t_1$ is the magnitude of the dominant nearest-neighbor hopping (cf.~Ref.~\citen{Raghu2008}).
For these choices of the order parameter, the gap magnitude in the spin-singlet case varies around $\approx0.75\Delta_0$ along the electron pockets and the inner hole pocket, and around $\approx0.6\Delta_0$ along the outer hole pocket, with opposite sign in the case of $s_\pm$~wave symmetry.
In the spin-triplet case instead, the gap magnitude varies around $\approx0.65\Delta_0$ along the electron pockets and the inner hole pocket, and around $\approx0.83\Delta_0$ along the outer hole pocket.
The inter-band interaction in the RPA is fixed to the value $U=W/4=3t_1$.

At first we study the general behavior of the spectra in a large energy range.
\Cref{fig:raghu-confront} shows the charge and spin DSF as a function of the transferred momentum $\bf q$ in an energy range up to $\hbar\omega\in[0,t]$, calculated in the RPA using \cref{eq:RPA}.
We consider the two cases of $s_\pm$ and $s_{++}$~wave symmetry.
A dispersive resonance peak is clearly visible at rather small momentum transfer, which arises due to the interaction processes.
Significant differences between charge and spin DSF are obtained at low-energy values in the order of the SC gap.
In particular, the spectral weight of the spin DSF at momentum vectors close to the points $(0,0)$ and $(\pi,\pi)$ is strongly suppressed, whereas spectral intensities at ${\bf Q}_{\rm AF}=(\pi,0)$ are suppressed in the charge (spin) spectra in the $s_\pm$ ($s_{++}$) wave state.
This spectral redistribution is indeed sensitive to the symmetry of the SC order parameter\cite{Marra2013}.
In order to investigate this feature in more detail, we will focus hereafter on the differences between charge and spin DSF by fixing the transferred energy $\hbar\omega$ to a value which is comparable with the energy scale of the SC gap.

\begin{figure}[t]
\centering\includegraphics[scale=1.55]{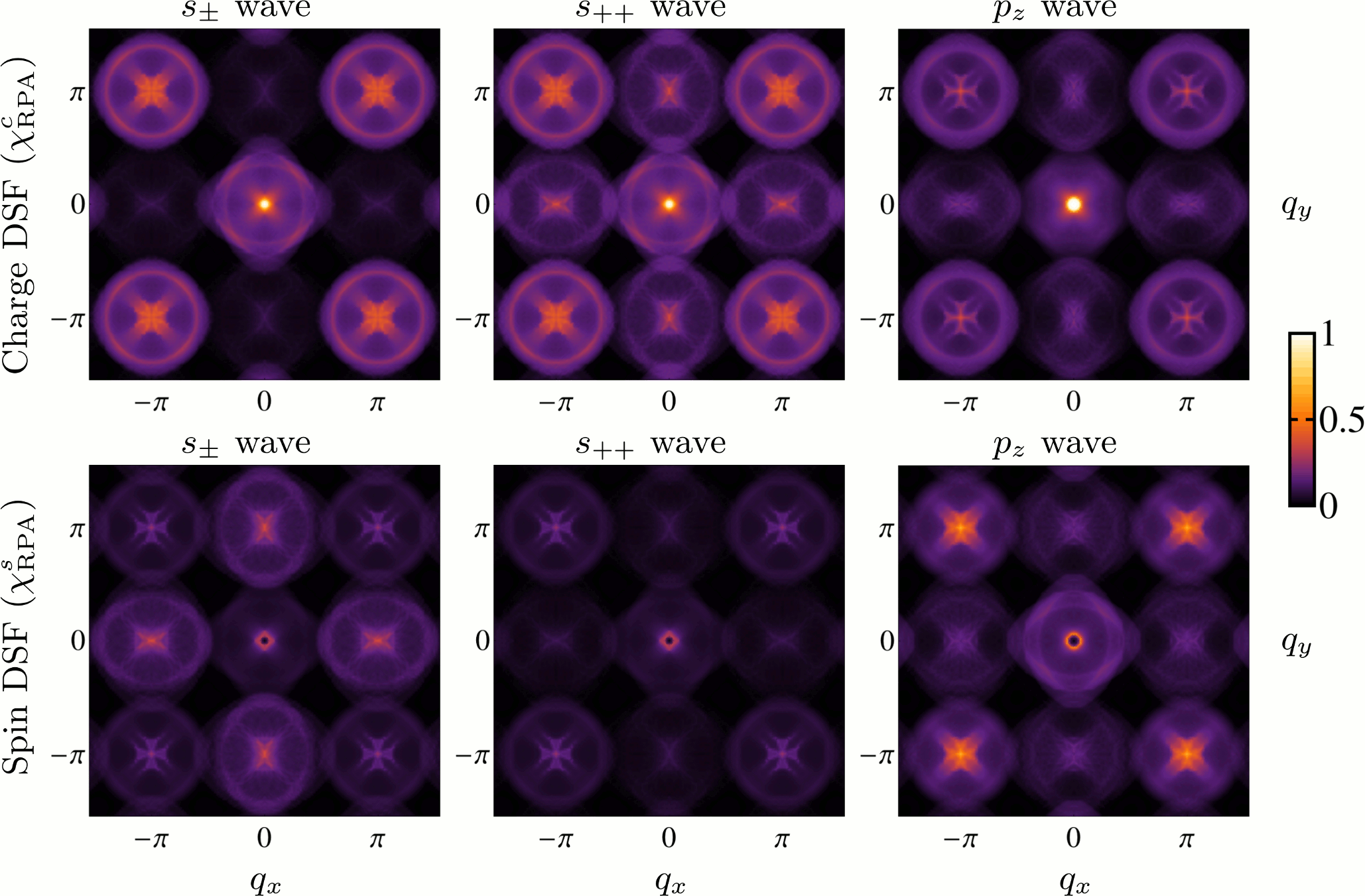}
\caption{
Charge and spin DSF $\chi^{c,s}_{\rm RPA}$ of quasiparticle excitations in LaOFeAs at a fixed energy loss $\hbar\omega=2\Delta_0$ as a function of the transferred momentum $\bf q$, with $s_\pm$, $s_{++}$, and $p_z$~wave order parameter ($\Delta_0=0.1t_1$, see main text), calculated using \cref{eq:StructureFactor,eq:RPA} assuming the bare electron dispersion and the orbital symmetry of the tight-binding model in Ref.~\citen{Raghu2008}, summing over inter-orbital and intra-orbital contributions.
The coherence peak from \cref{fig:raghu-confront} which appears here close to the point $(0,0)$ is largely dominant in the charge DSF spectra, while intensity distributions around ${\bf Q}_{\rm AF}=(\pi,0)$ and at $|{\bf q}|\approx\pi/2$ are sensitive to the differences in the order parameter phase along the Fermi surface.
Spectral intensities at ${\bf Q}_{\rm AF}$ are strongly suppressed in the charge (spin) spectra in the $s_\pm$ ($s_{++}$) wave state, while intensities in the region $|{\bf q}|\approx\pi/2$ around the point $(0,0)$ are suppressed in the charge (spin) DSF in the $p_z$ ($s_\pm$ or $s_{++}$) wave state.
Intensities are in arbitrary units.
}
\label{fig:raghu}
\end{figure}

For these purposes, we show in \cref{fig:raghu} the charge and spin DSF at a fixed energy loss $\hbar\omega=2\Delta_0$ as a function of the transferred momentum $\bf q$, for the three choices of the order parameter defined above.
As one can see, low-energy excitations which are sign-reversing, (opposite phase of the order parameter), suppress the charge component of the DSF, whereas sign-preserving excitations (same phase of the order parameter) suppress the spin component in the low-energy quasiparticle spectra.
For this reason, spectral intensities at ${\bf Q}_{\rm AF}=(\pi,0)$ in \cref{fig:raghu} are suppressed in the charge and in the spin DSF respectively in $s_\pm$~wave and in the $s_{++}$~wave SC states.
Such transferred momentum, which corresponds to the ordering vector of the antiferromagnetic phase, is in fact a nesting vector between the hole pockets and the electron pockets in the Brillouin zone, which have an opposite sign or the same sign of the order parameter alternatively in the $s_\pm$~wave and in the $s_{++}$~wave states.
Note that the enhancement of spectral intensity in the spin response functions at momentum ${\bf Q}_{\rm AF}$, as it appears in Fig.~\ref{fig:raghu} for the $s_\pm$ case, has been found also in INS experiments\cite{Qiu2008,Inosov2010}.

On the other hand, based on the result in \cref{fig:raghu} we propose that RIXS will be able to detect a characteristic signature of the $p$-wave order parameter.
Namely, the odd-symmetry in momentum space $\Delta_{\bf -k}=-\Delta_{\bf k}$ should produce signatures in the spectral intensities of excitations with transferred momentum $|{\bf q}|\approx\pi/2$ (see \cref{fig:raghu}), corresponding to a \emph{self-nesting} of the hole pockets.
This type of excitations, which lead to characteristic intensity features also in LiFeAs (see next Section), refer to \emph{intra-band} contributions located in a narrow momentum range similar to the conventional nesting scenario between the electron and hole pockets.
In the $s$-wave case these excitations preserve the sign of the order parameter ($\Delta_{\bf k+q}=\Delta_{\bf -k}=\Delta_{\bf k}$), leading to a suppression of spectral intensities in the spin DSF\@.
In the $p$-wave case instead, these excitations are sign-reversing ($\Delta_{\bf k+q}=\Delta_{\bf -k}=-\Delta_{\bf k}$), with a consequent suppression in the charge DSF\@.

\subsection*{LiFeAs}

While there is a general agreement about the presence of a spin-singlet $s_\pm$~wave superconductivity\cite{Mazin2008,Kuroki2008} in other iron-based SC, where nesting dominates the low-energy properties, the nature of the SC state in LiFeAs seems to be elusive.
Different scenarios have been proposed in place of the $s_\pm$~wave pairing, e.g., an $s_{++}$~wave SC state, driven by the critical $3d$-orbital fluctuations induced by moderate electron-phonon interactions\cite{Kontani2010}, or even a spin-triplet pairing driven by ferromagnetic fluctuations\cite{Brydon2011}.
While the singlet pairing is supported by some INS experiments\cite{Taylor2011}, the unusual shape of the Fermi surface and the momentum dependency of the SC gap measured by ARPES\cite{Borisenko2012} is in conflict with the $s_\pm$~wave symmetry.
Moreover, STM experiments of the quasiparticle interference\cite{Hanke2012} are consistent with a $p$-wave spin-triplet state or with a singlet pairing mechanism with a more complex order parameter ($s+\imath d$~wave).
Whereas ARPES has been proven to be powerful in measuring the momentum dependence of the SC gap on the Fermi surface\cite{Borisenko2010,Borisenko2012}, it should be noted here that ARPES, since not sensitive to the order parameter phase, cannot distinguish between singlet and triplet pairing, i.e., between even ($\Delta_{\bf k}=\Delta_{-\bf k}$) and odd ($\Delta_{\bf k}=-\Delta_{-\bf k}$) symmetry of the order parameter.
In fact, the experimental momentum dependence of the SC gap measured by ARPES\cite{Borisenko2012} is consistent, in principle, with a spin-singlet as well as with a spin-triplet state, as long as the pairing mechanism correctly reproduces the gap magnitude variations along the Fermi surface.

\begin{figure}[t]
\centering\includegraphics[scale=.2]{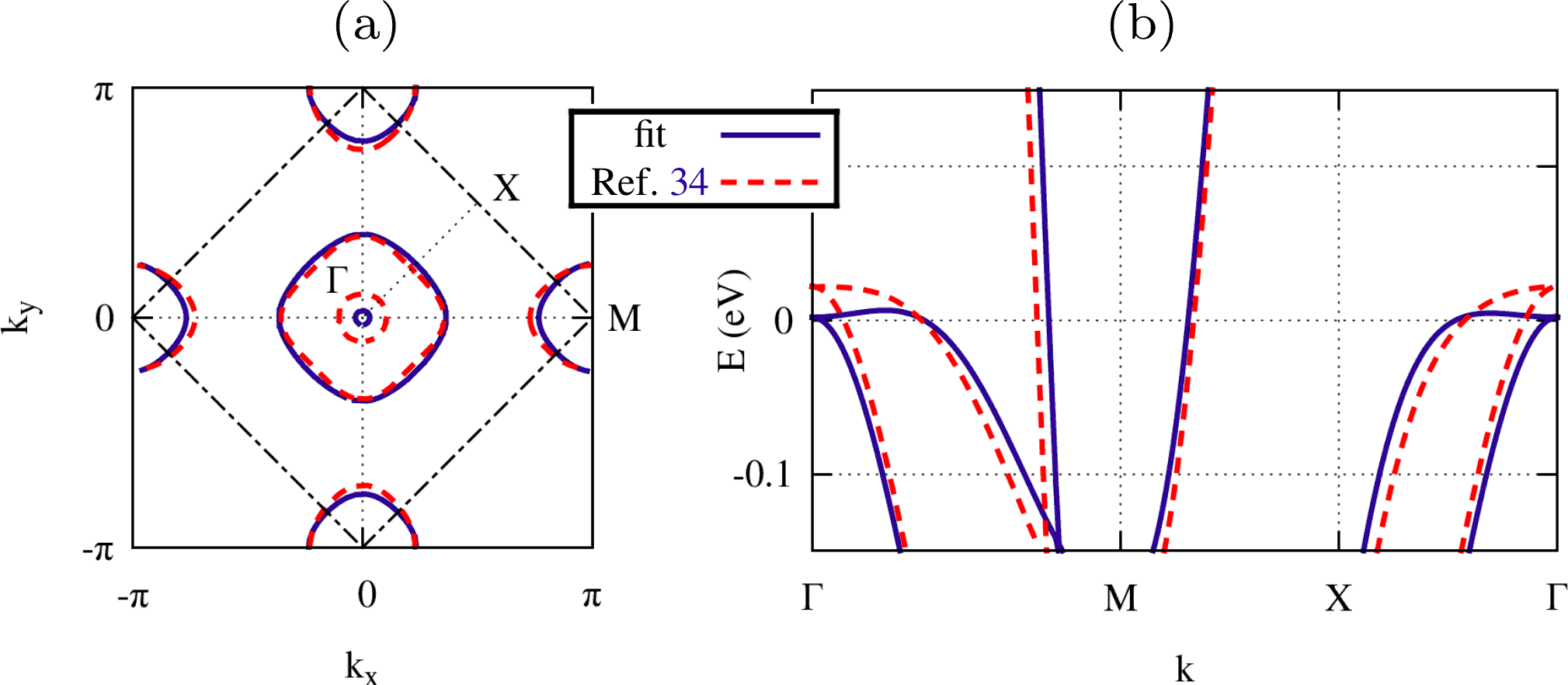}
\caption{
Fermi surface (a) and electronic dispersion (b) for the tight-binding model of Ref.~\citen{Brydon2011} (dashed line) and for the same model with hopping parameters (see \cref{tab:tb}) fitted with the experimental Fermi surface of LiFeAs\cite{Knolle2012,Borisenko2010} (solid line).
}
\label{fig:bands}
\end{figure}

The theoretical predictions shown in this Section in combination with an appropriate RIXS experiment might help to clarify the complicated and controversial situation of LiFeAs.
To achieve this goal, we consider different order parameter symmetries, corresponding to spin-singlet and spin-triplet pairing.
In order to properly take into account the orbital degrees of freedom of the system, we construct our model on the basis of the effective three-band tight-binding model proposed in Ref.~\citen{Brydon2011}, which includes the effective hoppings between the $t_{2g}$ orbitals of the iron ions, within a single Fe-ion unit cell.
Nevertheless, a comparison with ARPES measurements\cite{Knolle2012,Borisenko2010} shows that the inner hole pocket in LiFeAs is much smaller than the one produced by the tight-binding model in Ref.~\citen{Brydon2011}.
Furthermore, the SC gap is significantly larger\cite{Borisenko2012} on the inner hole pocket than on the outer one.
For this reason, we redefine the hopping parameters in order to fit the experimental Fermi surface\cite{Knolle2012,Borisenko2010}.
These parameters are given in \cref{tab:tb}, while in \cref{fig:bands} we compare the fitted Fermi surface (a) and bare electron dispersion (b) with the original model (cf.~Ref.~\citen{Brydon2011}).

\begin{table}[t]
\centering
\begin{tabular*}{\textwidth}{c @{\extracolsep{\fill}} cccccccccccccc}
		\hline\hline
		&& $t_1$ & $t_2$ & $t_3$ & $t_4$ & $t_5$ & $t_6$ & $t_7$
		& $t_8$ & $t_9$ & $t_{10}$ & $t_{11}$ & $\Delta_{xy}$ \\
		\hline
		fit &&
		$0.019$ & $0.123$ & $0.014$ & $-0.055$ & $0.217$ & $0.264$ & $-0.137$
		& $-t_7/2$ & $-0.060$ & $-0.057$ & $0.016$ & $1$\\
		Ref.~\citen{Brydon2011}&\qquad&
		$0.020$ & $0.120$ & $0.020$ & $-0.046$ & $0.200$ & $0.300$ & $-0.150$
		& $-t_7/2$ & $-0.060$ & $-0.030$ & $0.014$ & $1$\\
\end{tabular*}
\caption{
Hopping parameters for the three-band effective tight-binding model (see Ref.~\citen{Brydon2011} for the definitions of the parameters and for details) compared with those fitted with the experimental Fermi surface of LiFeAs\cite{Knolle2012,Borisenko2010}.
The chemical potential is $\mu=0.338$, that corresponds to a filling of four electrons per site.
Energy units are in electron volts.
}
\label{tab:tb}
\end{table}

\begin{figure}[t]
\centering\includegraphics[scale=1.55]{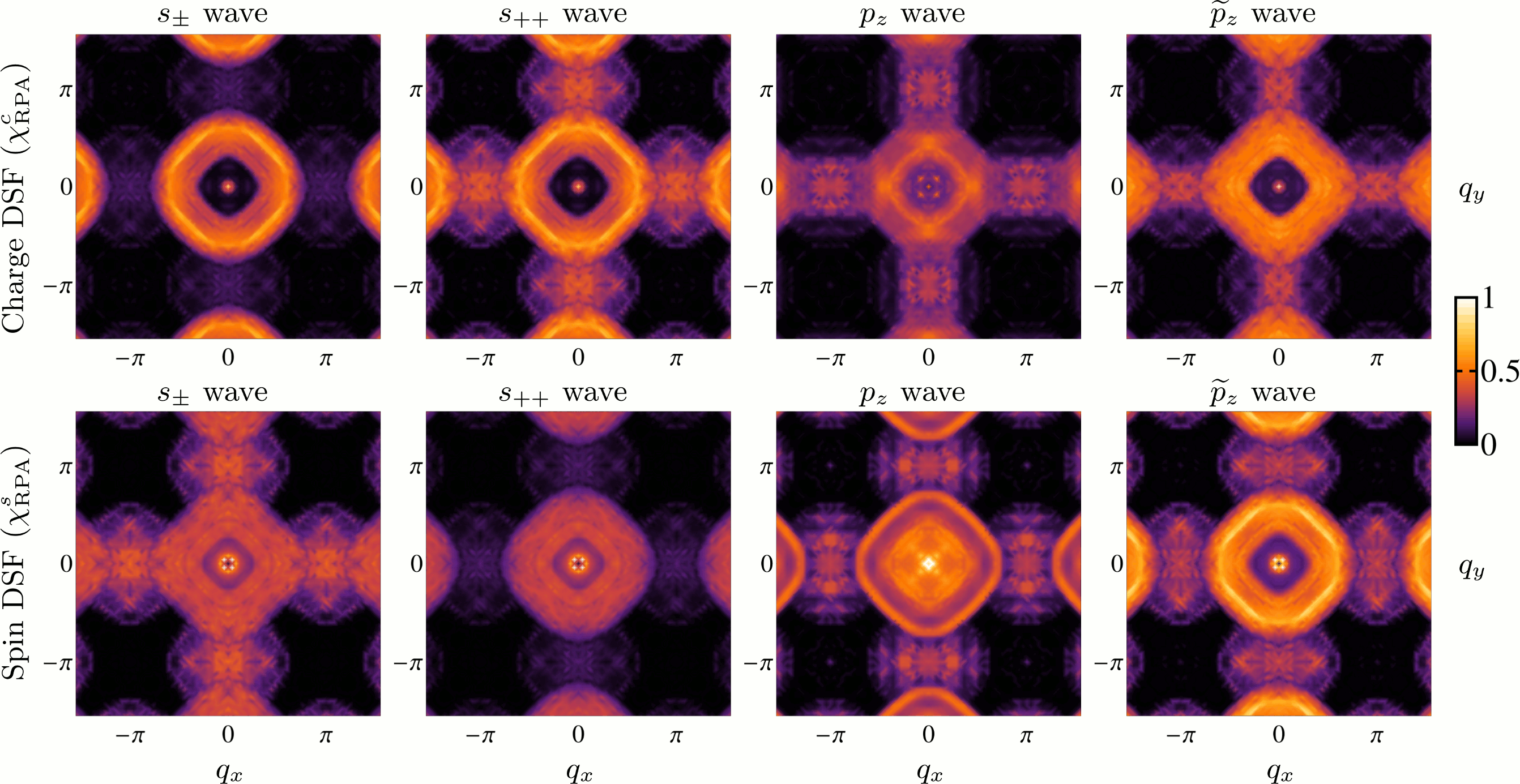}
\caption{
Charge and spin DSF $\chi^{c,s}_{\rm RPA}$ of quasiparticle excitations in LiFeAs at a fixed energy loss $\hbar\omega=2\Delta_0=12\meV$ as a function of the transferred momentum $\bf q$, with the $s_\pm$, $s_{++}$, $p_z$, and $\widetilde{p}_z$~wave SC order parameter, calculated using \cref{eq:StructureFactor,eq:RPA} assuming the bare electron dispersion and the orbital symmetry of the tight-binding mode in Ref.~\citen{Brydon2011} fitted with ARPES data\cite{Knolle2012,Borisenko2010}, and summing over inter-orbital and intra-orbital contributions.
Spectral intensities at ${\bf Q}_{\rm AF}=(\pi,0)$ are suppressed in the charge (spin) spectra in the $s_\pm$ ($s_{++}$) wave state, while intensities in the region $|{\bf q}|\approx\pi/2$ around the point $(0,0)$ are suppressed in the charge (spin) DSF in the $p_z$ and $\widetilde{p}_z$ ($s_\pm$ and $s_{++}$) state.
Intensities are in arbitrary units.
}
\label{fig:lifeas}
\end{figure}

Besides the $s_\pm$, $s_{++}$, and $p_z$~wave defined above, we consider here also a triplet pairing order parameter $\widetilde{p}_z$, defined as $\Delta^{\widetilde{p}_z}_{\bf k}=|\Delta^{s_{\pm}}_{\bf k}|e^{\imath\phi_{\bf k}}$, i.e., having the same magnitude of the $s_{\pm}$ (or $s_{++}$) wave and the same phase $\phi_{\bf k}=\arg{\Delta^{p_z}_{\bf k}}$ of the $p_z$~wave order parameter.
This SC order parameter is considered here for comparison, in order to have an example of a spin-triplet pairing which reproduces the experimental gap magnitude on the different branches of the Fermi surface in LiFeAs.
Moreover, the equal gap structure in comparison to the singlet pairing models allows us to study those features of spectra which are solely attributed to the phase variation.
We fix the order parameter magnitude by taking $\Delta_0=6\meV$, in order to be consistent with the measured value of the SC gap in LiFeAs\cite{Borisenko2012}.
We consider the inter-band interaction in the RPA as $U=W/4\approx0.7$~eV.
Therefore, in the case of the $s$-wave states ($s_\pm$ and $s_{++}$), and of the $\widetilde{p}_z$~wave state, the gap magnitude varies around $\approx4.6\meV$ along the electron pockets, around $\approx6\meV$ along the inner hole pocket, and around $\approx3\meV$ along the outer hole pocket, with opposite sign in the case of the $s_\pm$~wave symmetry, and with the phase continuously varying on the Fermi surface in the case of the $\widetilde{p}_z$~wave state.
In the $p_z$~wave case instead, the gap magnitude varies around $\approx4\meV$ along the electron pockets, around $\approx0.6\meV$ along the inner hole pocket, and around $\approx5.6\meV$ along the outer hole pocket.
In any of the case considered, the low-energy quasiparticle excitations contribute to coherence peaks at $(0,0)$ with energy in the range $6\meV<E<12\meV$ ($\Delta_0<E<2\Delta_0$).

In \cref{fig:lifeas} we show the RIXS intensities for the charge and spin DSF at a fixed energy loss $\hbar\omega=2\Delta_0=12 \meV$ as a function of the transferred momentum $\bf q$, for different choices of the SC order parameter symmetry, calculated using \cref{eq:StructureFactor}.
The resonant peaks at ${\bf Q}_{\rm AF}$ and at $|{\bf q}|\approx\pi/2$ are clearly visible in the calculated spectra, and are consistent with recent INS experiments\cite{Knolle2012,Qureshi2012}.
However, in LiFeAs, no nesting occurs between the hole and the electron pockets\cite{Borisenko2010}, and therefore the peak at ${\bf Q}_{\rm AF}$ in the quasiparticle spectra, which corresponds to the scattering between hole and electron pockets, is much weaker and broader than in the LaOFeAs case.
We find a square-like intensity distribution at small momenta for all the considered pairing symmetries, which is a typical feature of the low-energy spectrum in LiFeAs arising from inter-band scattering processes between the two hole-like Fermi surface branches\cite{Hanke2012,Hess2013}.
Indeed, as in the previous case, RIXS spectra in LiFeAs are strongly sensitive to the symmetry of the SC order parameter and on its relative phase differences along the Fermi surface.
In fact, spectral intensities at ${\bf Q}_{\rm AF}$ are further suppressed in the charge and in the spin DSF respectively in the $s_\pm$~wave and in the $s_{++}$~wave SC states.
This is because one has $\Delta_{{\bf k}+{\bf Q}_{\rm AF}}=\pm\Delta_{\bf k}$, with the $\pm$ sign corresponding to the $s_{++}$ and $s_\pm$~wave, resulting in sign-preserving and sign-reversing excitations respectively.
In the $p$-wave states no suppression occurs, being $\Delta_{{\bf k}+{\bf Q}_{\rm AF}}=\Delta_{\bf k}^*$, i.e., with a phase difference given by $2\phi_{\bf k}$, resulting in charge and spin coherence factors [see \cref{eq:CoherenceFactors}] which continuously vary on the Fermi surface.
Again, the signature of the $p$-wave odd-symmetry is in the spectral intensities of excitations with transferred momentum $|{\bf q}|\approx\pi/2$, corresponding to a self-nesting of the larger hole pocket (see \cref{fig:lifeas}).
While in the $s$-wave case excitations with $|{\bf q}|\approx\pi/2$ are sign-preserving, with a consequent suppression of spectral intensities in the spin DSF, in the $p$-wave case they are sign-reversing, resulting instead in an enhancement in the spin DSF\@.

\begin{figure}[t]
\centering\includegraphics[scale=1.45]{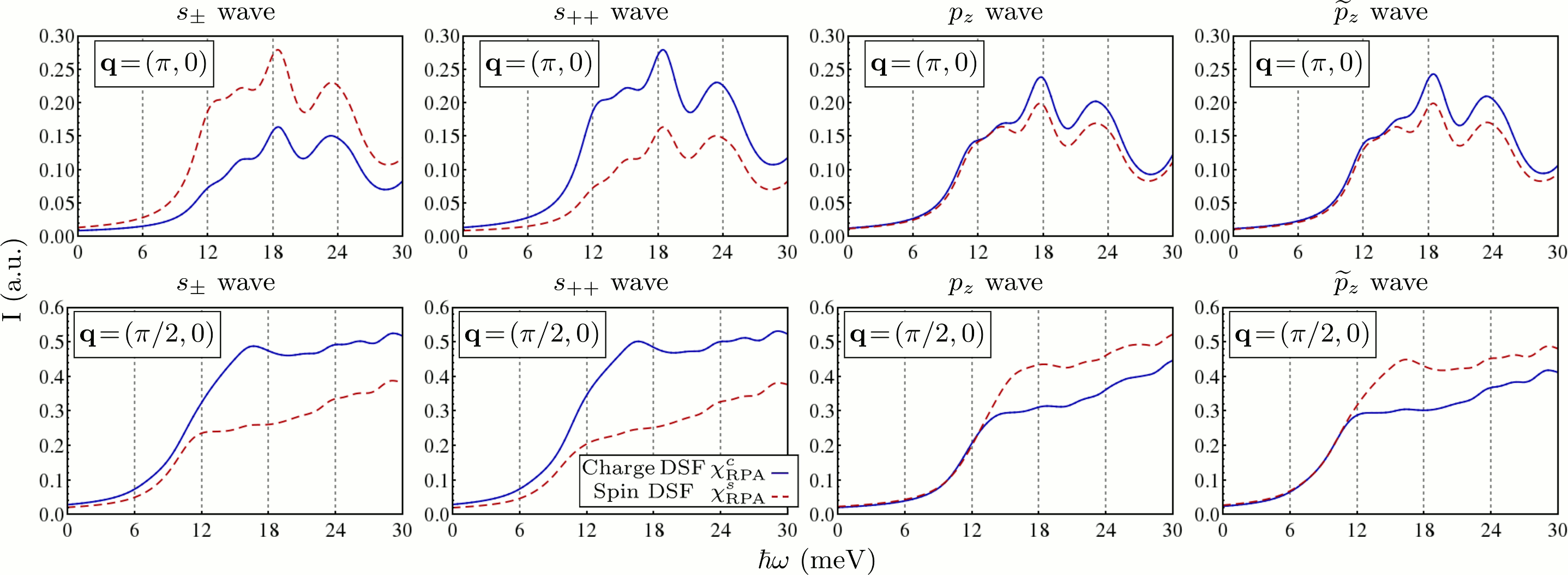}
\caption{
Charge (solid line) and spin (dashed line) DSF $\chi^{c,s}_{\rm RPA}$ of quasiparticle excitations in LiFeAs at ${\bf Q}_{\rm AF}=(\pi,0)$ and at $(\pi/2,0)$ as a function of the energy loss, calculated using \cref{eq:StructureFactor,eq:RPA}, respectively with $s_\pm$, $s_{++}$, $p_z$, and $\widetilde{p}_z$~wave SC order parameter ($\Delta_0=6\meV$).
Spectral intensities at ${\bf Q}_{\rm AF}$ are larger for the spin (charge) DSF spectra in the $s_\pm$ ($s_{++}$) wave state, while intensities at $(\pi/2,0)$ are larger in the spin (charge) DSF in the $p_z$ and $\widetilde{p}_z$ ($s_\pm$ and $s_{++}$) wave state.
Intensities are in arbitrary units.
}
\label{fig:lifeas-hsp}
\end{figure}

In order to present in the most clear way how to distinguish between the different pairing scenarios in LiFeAs, we show in \cref{fig:lifeas-hsp} the RIXS spectra as a function of the energy loss for the charge and spin DSF of quasiparticle excitations at ${\bf Q}_{\rm AF}$ and at $(\pi/2)$, again for different choices of the SC order parameter symmetry.
As we have seen, these particular momenta are those where the sensitivity to the order parameter phase is more pronounced.
In particular, spectral intensities corresponding to the transferred momentum ${\bf Q}_{\rm AF}$ are sensitive to sign changes of the order parameter between hole and electron pockets.
Indeed, as one can see in \cref{fig:lifeas-hsp}, the charge (spin) DSF is suppressed in the $s_\pm$ ($s_{++}$) wave state.
Therefore, a comparison between charge and spin DSF can be revealing of a sign-reversal in the order parameter between disconnected branches of the Fermi surface.
On the other hand, the spectral contributions of the intra-band scattering within the hole pockets, which correspond to a transferred momentum $|{\bf q}|\approx\pi/2$, are strongly affected by the parity of the order parameter, and therefore can discriminate between spin-singlet (e.g., $s$-wave) and spin-triplet pairing (e.g., $p$-wave).
In fact, spectral intensities at $(\pi/2,0)$ in \cref{fig:lifeas-hsp} are suppressed in the charge and spin DSF respectively in the spin-triplet ($p_z$ and $\widetilde{p}_z$~wave) and in the spin-singlet ($s_\pm$ and $s_{++}$~wave) cases.
It should be noticed here that this result is general, and does not depend on the gap magnitude dependence along the Fermi surface, but only on its phase variations, and therefore is a mere consequence of the odd parity of the order parameter.
This aspect is clearly displayed by the two panels of \cref{fig:lifeas-hsp} referring to the $p$-wave state at $(\pi/2,0)$.
The suppression of the charge DSF occurs for both $p_z$ and $\widetilde{p}_z$~wave, which have a different gap magnitude dependence, but nevertheless the same phase variations and the same parity.

\section*{Conclusions}

We have shown that RIXS spectra of quasiparticle excitations are sensitive to phase differences of the SC order parameter along the Fermi surface, and hence allow one to distinguish among different SC states, in particular between spin-singlet and spin-triplet pairing and between sign-preserving and sign-reversing $s$-wave states in iron-based superconductors.
In particular, RIXS spectral intensities corresponding to a self-nesting of the hole pockets can discriminate between singlet and triplet pairing, while RIXS spectra corresponding to a scattering between hole and electron pockets [${\bf Q}_{\rm AF}=(\pi,0)$] can discriminate between an $s_\pm$~wave and an $s_{++}$~wave order parameter.
The contribution of quasiparticle excitations can be separated from other effects (e.g., fluorescence) by considering the difference between the total inelastic scattering measured slightly above and slightly below the critical temperature.
Therefore RIXS has the potential to serve as a tool to probe the symmetry of the SC order parameter in iron-based superconductors, as soon as the energy resolution will reach the energy scale of the SC gap.

\section*{Acknowledgements}

We thank Bernd B\"uchner for useful discussions.
This research acknowledges support from the Computational Materials Science Network (CMSN) program of the Division of Materials Science and Engineering, U.\,S.\ Department of Energy, and from the SPP 1458 program of the German Research Foundation (Deutsche Forschungsgemeinschaft, DFG).

\section*{Author contributions}

P.\,M.\ and S.\,S.\ have performed the calculations.
The results were discussed and reviewed by all authors.
All authors contributed to the main text.
P.\,M.\ prepared the figures.

\section*{Additional information}
\textbf{Competing financial interests:} The authors declare no competing financial interests.
\\\quad\\
\textbf{How to cite this article:} Marra, P. \emph{et al.} Theoretical approach to resonant inelastic x-ray scattering in iron-based superconductors at the energy scale of the superconducting gap. \emph{Sci. Rep.} \textbf{6}, 25386; doi: \href{http://dx.doi.org/10.1038/srep25386}{10.1038/srep25386} (2016).
\\\quad\\
This work is licensed under a Creative Commons Attribution 4.0 International License. The images or other third party material in this article are included in the article’s Creative Commons license, unless indicated otherwise in the credit line; if the material is not included under the Creative Commons license, users will need to obtain permission from the license holder to reproduce the material. To view a copy of this license, visit \href{http://creativecommons.org/licenses/by/4.0/}{http://creativecommons.org/licenses/by/4.0/}


\begin{thebibliography}{10}

\bibitem{Leggett2006}
A.~J. Leggett,
{\em What DO we know about high $T_c$?},
\href{http://dx.doi.org/10.1038/nphys254}{Nat. Phys. {\bf 2}, 134 } (2006).

\bibitem{Kirtley1995}
J.~R. Kirtley, C.~C. Tsuei, J.~Z. Sun, C.~C. Chi, L.~S. Yu-Jahnes, A. Gupta, M. Rupp, and M.~B. Ketchen,
{\em Symmetry of the order parameter in the high-$T_c$ superconductor YBa$_2$Cu$_3$O$_{7-\delta}$},
\href{http://dx.doi.org/10.1038/373225a0}{Nature {\bf 373}, 225 } (1995).

\bibitem{Tsuei1997}
C.~C. Tsuei, J.~R. Kirtley, Z.~F. Ren, J.~H. Wang, H. Raffy, and Z.~Z. Li,
{\em Pure $d_{x^2 - y^2}$ order-parameter symmetry in the tetragonal superconductor TI$_2$Ba$_2$CuO$_{6+\delta}$},
\href{http://dx.doi.org/10.1038/387481a0}{Nature {\bf 387}, 481 } (1997).

\bibitem{Damascelli2003}
A. Damascelli, Z. Hussain, and Z.-X. Shen,
{\em Angle-resolved photoemission studies of the cuprate superconductors},
\href{http://dx.doi.org/10.1103/RevModPhys.75.473}{Rev. Mod. Phys. {\bf 75}, 473 } (2003).

\bibitem{Kuroki2001}
K. Kuroki and R. Arita,
{\em Possible high-${T}_{c}$ superconductivity mediated by antiferromagnetic spin fluctuations in systems with Fermi surface pockets},
\href{http://dx.doi.org/10.1103/PhysRevB.64.024501}{Phys. Rev. B {\bf 64}, 024501 } (2001).

\bibitem{Mazin2008}
I.~I. Mazin, D.~J. Singh, M.~D. Johannes, and M.~H. Du,
{\em Unconventional superconductivity with a sign reversal in the order parameter of LaFeAsO$_{1-x}$F$_{x}$},
\href{http://dx.doi.org/10.1103/PhysRevLett.101.057003}{Phys. Rev. Lett. {\bf 101}, 057003 } (2008).

\bibitem{Kuroki2008}
K. Kuroki, S. Onari, R. Arita, H. Usui, Y. Tanaka, H. Kontani, and H. Aoki,
{\em Unconventional pairing originating from the disconnected Fermi surfaces of superconducting LaFeAsO$_{1-x}$F$_{x}$},
\href{http://dx.doi.org/10.1103/PhysRevLett.101.087004}{Phys. Rev. Lett. {\bf 101}, 087004 } (2008).

\bibitem{Paglione2010}
J. Paglione and R.~L. Greene,
{\em High-temperature superconductivity in iron-based materials},
\href{http://dx.doi.org/10.1038/nphys1759}{Nat. Phys. {\bf 6}, 645 } (2010).

\bibitem{Stewart2011}
G.~R. Stewart,
{\em Superconductivity in iron compounds},
\href{http://dx.doi.org/10.1103/RevModPhys.83.1589}{Rev. Mod. Phys. {\bf 83}, 1589 } (2011).

\bibitem{Chubukov2012}
A. Chubukov,
{\em Pairing mechanism in Fe-based superconductors},
\href{http://dx.doi.org/10.1146/annurev-conmatphys-020911-125055}{Annu. Rev. Condens. Matter Phys. {\bf 3}, 57 } (2012).

\bibitem{Hosono2015}
H. Hosono and K. Kuroki,
{\em Iron-based superconductors: Current status of materials and pairing mechanism},
\href{http://dx.doi.org/10.1016/j.physc.2015.02.020}{Phys. C. {\bf 514}, 399 } (2015).

\bibitem{Tsuei2000}
C.~C. Tsuei and J.~R. Kirtley,
{\em Pairing symmetry in cuprate superconductors},
\href{http://dx.doi.org/10.1103/RevModPhys.72.969}{Rev. Mod. Phys. {\bf 72}, 969 } (2000).

\bibitem{Chen2010}
C.-T. Chen, C.~C. Tsuei, M.~B. Ketchen, Z.-A. Ren, and Z.~X. Zhao,
{\em Integer and half-integer flux-quantum transitions in a niobium-iron pnictide loop},
\href{http://dx.doi.org/10.1038/nphys1531}{Nat. Phys. {\bf 6}, 260 } (2010).

\bibitem{Hoffman2002}
J.~E. Hoffman, K. McElroy, D.-H. Lee, K.~M. Lang, H. Eisaki, S. Uchida, and J.~C. Davis,
{\em Imaging quasiparticle interference in Bi$_2$Sr$_2$CaCu$_2$O$_{8+\delta}$},
\href{http://dx.doi.org/10.1126/science.1072640}{Science {\bf 297}, 1148 } (2002).

\bibitem{McElroy2003}
K. McElroy, R.~W. Simmonds, J.~E. Hoffman, D.-H. Lee, J. Orenstein, H. Eisaki, S. Uchida, and J.~C. Davis,
{\em Relating atomic-scale electronic phenomena to wave-like quasiparticle states in superconducting Bi$_2$Sr$_2$CaCu$_2$O$_{8+\delta}$},
\href{http://dx.doi.org/10.1038/nature01496}{Nature {\bf 422}, 592 } (2003).

\bibitem{Hanaguri2007}
T. Hanaguri, Y. Kohsaka, J.~C. Davis, C. Lupien, I. Yamada, M. Azuma, M. Takano, K. Ohishi, M. Ono, and H. Takagi,
{\em Quasiparticle interference and superconducting gap in Ca$_{2-x}$Na$_x$CuO$_2$Cl$_2$},
\href{http://dx.doi.org/10.1038/nphys753}{Nat. Phys. {\bf 3}, 865 } (2007).

\bibitem{Kohsaka2008}
Y. Kohsaka, C. Taylor, P. Wahl, A. Schmidt, J. Lee, K. Fujita, J.~W. Alldredge, K. McElroy, J. Lee, H. Eisaki, S. Uchida, D.-H. Lee, and J.~C. Davis,
{\em How Cooper pairs vanish approaching the Mott insulator in Bi$_2$Sr$_2$CaCu$_2$O$_{8+\delta}$},
\href{http://dx.doi.org/10.1038/nature07243}{Nature {\bf 454}, 1072 } (2008).

\bibitem{Hanaguri2009}
T. Hanaguri, Y. Kohsaka, M. Ono, M. Maltseva, P. Coleman, I. Yamada, M. Azuma, M. Takano, K. Ohishi, and H. Takagi,
{\em Coherence factors in a high-T$_c$ cuprate probed by quasi-particle scattering off vortices},
\href{http://dx.doi.org/10.1126/science.1166138}{Science {\bf 323}, 923 } (2009).

\bibitem{Maier2008}
T.~A. Maier and D.~J. Scalapino,
{\em Theory of neutron scattering as a probe of the superconducting gap in the iron pnictides},
\href{http://dx.doi.org/10.1103/PhysRevB.78.020514}{Phys. Rev. B {\bf 78}, 020514 } (2008).

\bibitem{Korshunov2008}
M.~M. Korshunov and I. Eremin,
{\em Theory of magnetic excitations in iron-based layered superconductors},
\href{http://dx.doi.org/10.1103/PhysRevB.78.140509}{Phys. Rev. B {\bf 78}, 140509 } (2008).

\bibitem{Christianson2008}
A.~D. Christianson, E.~A. Goremychkin, R. Osborn, S. Rosenkranz, M.~D. Lumsden, C.~D. Malliakas, I.~S. Todorov, H. Claus, D.~Y. Chung, M.~G. Kanatzidis, R.~I. Bewley, and T. Guidi,
{\em Unconventional superconductivity in Ba$_{0.6}$K$_{0.4}$Fe$_2$As$_2$ from inelastic neutron scattering},
\href{http://dx.doi.org/10.1038/nature07625}{Nature {\bf 456}, 930 } (2008).

\bibitem{Maier2011}
T.~A. Maier, S. Graser, P.~J. Hirschfeld, and D.~J. Scalapino,
{\em Inelastic neutron and x-ray scattering as probes of the sign structure of the superconducting gap in iron pnictides},
\href{http://dx.doi.org/10.1103/PhysRevB.83.220505}{Phys. Rev. B {\bf 83}, 220505 } (2011).

\bibitem{Qiu2008}
Y. Qiu, M. Kofu, W. Bao, S.-H. Lee, Q. Huang, T. Yildirim, J.~R.~D. Copley, J.~W. Lynn, T. Wu, G. Wu, and X.~H. Chen,
{\em Neutron-scattering study of the oxypnictide superconductor LaFeAsO$_{0.87}$F$_{0.13}$},
\href{http://dx.doi.org/10.1103/PhysRevB.78.052508}{Phys. Rev. B {\bf 78}, 052508 } (2008).

\bibitem{Inosov2010}
D.~S. Inosov, J.~T. Park, P. Bourges, D.~L. Sun, Y. Sidis, A. Schneidewind, K. Hradil, D. Haug, C.~T. Lin, B. Keimer, and V. Hinkov,
{\em Normal-state spin dynamics and temperature-dependent spin-resonance energy in optimally doped BaFe$_{1.85}$Co$_{0.15}$As$_2$},
\href{http://dx.doi.org/10.1038/nphys1483}{Nat. Phys. {\bf 6}, 178 } (2010).

\bibitem{Knolle2012}
J. Knolle, V.~B. Zabolotnyy, I. Eremin, S.~V. Borisenko, N. Qureshi, M. Braden, D.~V. Evtushinsky, T.~K. Kim, A.~A. Kordyuk, S. Sykora, C. Hess, I.~V. Morozov, S. Wurmehl, R. Moessner, and B. B\"uchner,
{\em Incommensurate magnetic fluctuations and Fermi surface topology in LiFeAs},
\href{http://dx.doi.org/10.1103/PhysRevB.86.174519}{Phys. Rev. B {\bf 86}, 174519 } (2012).

\bibitem{Wang2008}
X. Wang, Q. Liu, Y. Lv, W. Gao, L. Yang, R. Yu, F. Li, and C. Jin,
{\em The superconductivity at 18~K in LiFeAs system},
\href{http://dx.doi.org/10.1016/j.ssc.2008.09.057}{Solid State Commun. {\bf 148}, 538 } (2008).

\bibitem{Tapp2008}
J.~H. Tapp, Z. Tang, B. Lv, K. Sasmal, B. Lorenz, P.~C.~W. Chu, and A.~M. Guloy,
{\em LiFeAs: An intrinsic FeAs-based superconductor with ${T}_{c}=18$K},
\href{http://dx.doi.org/10.1103/PhysRevB.78.060505}{Phys. Rev. B {\bf 78}, 060505 } (2008).

\bibitem{Borisenko2010}
S.~V. Borisenko, V.~B. Zabolotnyy, D.~V. Evtushinsky, T.~K. Kim, I.~V. Morozov, A.~N. Yaresko, A.~A. Kordyuk, G. Behr, A. Vasiliev, R. Follath, and B. B\"uchner,
{\em Superconductivity without nesting in LiFeAs},
\href{http://dx.doi.org/10.1103/PhysRevLett.105.067002}{Phys. Rev. Lett. {\bf 105}, 067002 } (2010).

\bibitem{Borisenko2012}
S.~V. Borisenko, V.~B. Zabolotnyy, A.~A. Kordyuk, D.~V. Evtushinsky, T.~K. Kim, I.~V. Morozov, R. Follath, and B. B\"uchner,
{\em One-sign order parameter in iron based superconductor},
\href{http://dx.doi.org/10.3390/sym4010251}{Symmetry {\bf 4}, 251 } (2012).

\bibitem{Platt2011}
C. Platt, R. Thomale, and W. Hanke,
{\em Superconducting state of the iron pnictide LiFeAs: A combined density-functional and functional-renormalization-group study},
\href{http://dx.doi.org/10.1103/PhysRevB.84.235121}{Phys. Rev. B {\bf 84}, 235121 } (2011).

\bibitem{Wang2013}
Y. Wang, A. Kreisel, V.~B. Zabolotnyy, S.~V. Borisenko, B. B\"uchner, T.~A. Maier, P.~J. Hirschfeld, and D.~J. Scalapino,
{\em Superconducting gap in LiFeAs from three-dimensional spin-fluctuation pairing calculations},
\href{http://dx.doi.org/10.1103/PhysRevB.88.174516}{Phys. Rev. B {\bf 88}, 174516 } (2013).

\bibitem{Ahn2014}
F. Ahn, I. Eremin, J. Knolle, V.~B. Zabolotnyy, S.~V. Borisenko, B. B\"uchner, and A.~V. Chubukov,
{\em Superconductivity from repulsion in LiFeAs: Novel $s$-wave symmetry and potential time-reversal symmetry breaking},
\href{http://dx.doi.org/10.1103/PhysRevB.89.144513}{Phys. Rev. B {\bf 89}, 144513 } (2014).

\bibitem{Kontani2010}
H. Kontani and S. Onari,
{\em Orbital-fluctuation-mediated superconductivity in iron pnictides: Analysis of the five-orbital Hubbard-Holstein model},
\href{http://dx.doi.org/10.1103/PhysRevLett.104.157001}{Phys. Rev. Lett. {\bf 104}, 157001 } (2010).

\bibitem{Brydon2011}
P.~M.~R. Brydon, M. Daghofer, C. Timm, and J. van~den Brink,
{\em Theory of magnetism and triplet superconductivity in LiFeAs},
\href{http://dx.doi.org/10.1103/PhysRevB.83.060501}{Phys. Rev. B {\bf 83}, 060501 } (2011).

\bibitem{Taylor2011}
A.~E. Taylor, M.~J. Pitcher, R.~A. Ewings, T.~G. Perring, S.~J. Clarke, and A.~T. Boothroyd,
{\em Antiferromagnetic spin fluctuations in LiFeAs observed by neutron scattering},
\href{http://dx.doi.org/10.1103/PhysRevB.83.220514}{Phys. Rev. B {\bf 83}, 220514 } (2011).

\bibitem{Hanke2012}
T. H\"anke, S. Sykora, R. Schlegel, D. Baumann, L. Harnagea, S. Wurmehl, M. Daghofer, B. B\"uchner, J. van~den Brink, and C. Hess,
{\em Probing the unconventional superconducting state of LiFeAs by quasiparticle interference},
\href{http://dx.doi.org/10.1103/PhysRevLett.108.127001}{Phys. Rev. Lett. {\bf 108}, 127001 } (2012).

\bibitem{Tinkham}
M. Tinkham,
{\em Introduction to superconductivity} (Dover Publications, 2004).

\bibitem{Hanaguri2010}
T. Hanaguri, S. Niitaka, K. Kuroki, and H. Takagi,
{\em Unconventional s-wave superconductivity in Fe(Se,Te)},
\href{http://dx.doi.org/10.1126/science.1187399}{Science {\bf 328}, 474 } (2010).

\bibitem{CommentHanaguri}
I.~I. Mazin and D.~J. Singh,
{\em Comment on ``Unconventional s-wave superconductivity in Fe(Se,Te)''},
\href{http://arxiv.org/abs/1007.0047}{arXiv:1007.0047} (2010).

\bibitem{ReplyCommentHanaguri}
T. Hanaguri, S. Niitaka, K. Kuroki, and H. Takagi,
{\em Reply to Comment on ``Unconventional s-wave superconductivity in Fe(Se,Te)''},
\href{http://arxiv.org/abs/1007.0307}{arXiv:1007.0307} (2010).

\bibitem{Sykora2011}
S. Sykora and P. Coleman,
{\em Quasiparticle interference in an iron-based superconductor},
\href{http://dx.doi.org/10.1103/PhysRevB.84.054501}{Phys. Rev. B {\bf 84}, 054501 } (2011).

\bibitem{Marra2013}
P. Marra, S. Sykora, K. Wohlfeld, and J. van~den Brink,
{\em Resonant inelastic x-ray scattering as a probe of the phase and excitations of the order parameter of superconductors},
\href{http://dx.doi.org/10.1103/PhysRevLett.110.117005}{Phys. Rev. Lett. {\bf 110}, 117005 } (2013).

\bibitem{Braicovich2009}
L. Braicovich, L.~J.~P. Ament, V. Bisogni, F. Forte, C. Aruta, G. Balestrino, N.~B. Brookes, G.~M. De~Luca, P.~G. Medaglia, F.~M. Granozio, M. Radovic, M. Salluzzo, J. van~den Brink, and G. Ghiringhelli,
{\em Dispersion of magnetic excitations in the cuprate La$_2$CuO$_4$ and CaCuO$_2$ compounds measured using resonant x-ray scattering},
\href{http://dx.doi.org/10.1103/PhysRevLett.102.167401}{Phys. Rev. Lett. {\bf 102}, 167401 } (2009).

\bibitem{Ulrich2009}
C. Ulrich, L.~J.~P. Ament, G. Ghiringhelli, L. Braicovich, M. Moretti~Sala, N. Pezzotta, T. Schmitt, G. Khaliullin, J. van~den Brink, H. Roth, T. Lorenz, and B. Keimer,
{\em Momentum dependence of orbital excitations in Mott-insulating titanates},
\href{http://dx.doi.org/10.1103/PhysRevLett.103.107205}{Phys. Rev. Lett. {\bf 103}, 107205 } (2009).

\bibitem{Yavas2010}
H. Yava\c{s}, M. van Veenendaal, J. van~den Brink, L.~J.~P. Ament, A. Alatas, B.~M. Leu, M.-O. Apostu, N. Wizent, G. Behr, W. Sturhahn, H. Sinn, and E.~E. Alp,
{\em Observation of phonons with resonant inelastic x-ray scattering},
\href{http://dx.doi.org/10.1088/0953-8984/22/48/485601}{J. Phys. Condens. Matter {\bf 22}, 485601 } (2010).

\bibitem{Ament2011}
L.~J.~P. Ament, M. van Veenendaal, T.~P. Devereaux, J.~P. Hill, and J. van~den Brink,
{\em Resonant inelastic x-ray scattering studies of elementary excitations},
\href{http://dx.doi.org/10.1103/RevModPhys.83.705}{Rev. Mod. Phys. {\bf 83}, 705 } (2011).

\bibitem{Ament2009}
L.~J.~P. Ament, G. Ghiringhelli, M.~M. Sala, L. Braicovich, and J. van~den Brink,
{\em Theoretical demonstration of how the dispersion of magnetic excitations in cuprate compounds can be determined using resonant inelastic x-ray scattering},
\href{http://dx.doi.org/10.1103/PhysRevLett.103.117003}{Phys. Rev. Lett. {\bf 103}, 117003 } (2009).

\bibitem{Kaneshita2011}
E. Kaneshita, K. Tsutsui, and T. Tohyama,
{\em Spin and orbital characters of excitations in iron arsenide superconductors revealed by simulated resonant inelastic x-ray scattering},
\href{http://dx.doi.org/10.1103/PhysRevB.84.020511}{Phys. Rev. B {\bf 84}, 020511 } (2011).

\bibitem{Zhou2013}
K.-J. Zhou, Y.-B. Huang, C. Monney, X. Dai, V.~N. Strocov, N.-L. Wang, Z.-G. Chen, C. Zhang, P. Dai, L. Patthey, J. van~den Brink, H. Ding, and T. Schmitt,
{\em Persistent high-energy spin excitations in iron-pnictide superconductors},
\href{http://dx.doi.org/10.1038/ncomms2428}{Nat. Commun. {\bf 4}, 1470 } (2013).

\bibitem{Harriger2011}
L.~W. Harriger, H.~Q. Luo, M.~S. Liu, C. Frost, J.~P. Hu, M.~R. Norman, and P. Dai,
{\em Nematic spin fluid in the tetragonal phase of BaFe${}_{2}$As${}_{2}$},
\href{http://dx.doi.org/10.1103/PhysRevB.84.054544}{Phys. Rev. B {\bf 84}, 054544 } (2011).

\bibitem{Hancock2010}
J.~N. Hancock, R. Viennois, D. van~der Marel, H.~M. R\o{}nnow, M. Guarise, P.-H. Lin, M. Grioni, M. Moretti~Sala, G. Ghiringhelli, V.~N. Strocov, J. Schlappa, and T. Schmitt,
{\em Evidence for core-hole-mediated inelastic x-ray scattering from metallic Fe$_{1.087}$Te},
\href{http://dx.doi.org/10.1103/PhysRevB.82.020513}{Phys. Rev. B {\bf 82}, 020513 } (2010).

\bibitem{Yang2009}
W.~L. Yang, A.~P. Sorini, C.-C. Chen, B. Moritz, W.-S. Lee, F. Vernay, P. Olalde-Velasco, J.~D. Denlinger, B. Delley, J.-H. Chu, J.~G. Analytis, I.~R. Fisher, Z.~A. Ren, J. Yang, W. Lu, Z.~X. Zhao, J. van~den Brink, Z. Hussain, Z.-X. Shen, and T.~P. Devereaux,
{\em Evidence for weak electronic correlations in iron pnictides},
\href{http://dx.doi.org/10.1103/PhysRevB.80.014508}{Phys. Rev. B {\bf 80}, 014508 } (2009).

\bibitem{Haverkort2010}
M.~W. Haverkort,
{\em Theory of resonant inelastic x-ray scattering by collective magnetic excitations},
\href{http://dx.doi.org/10.1103/PhysRevLett.105.167404}{Phys. Rev. Lett. {\bf 105}, 167404 } (2010).

\bibitem{Marra2012}
P. Marra, K. Wohlfeld, and J. van~den Brink,
{\em Unraveling orbital correlations with magnetic resonant inelastic x-ray scattering},
\href{http://dx.doi.org/10.1103/PhysRevLett.109.117401}{Phys. Rev. Lett. {\bf 109}, 117401 } (2012).

\bibitem{Andersen2005}
B.~M. Andersen and P. Hedeg\aa{}rd,
{\em Spin dynamics in the stripe phase of the cuprate superconductors},
\href{http://dx.doi.org/10.1103/PhysRevLett.95.037002}{Phys. Rev. Lett. {\bf 95}, 037002 } (2005).

\bibitem{Raghu2008}
S. Raghu, X.-L. Qi, C.-X. Liu, D.~J. Scalapino, and S.-C. Zhang,
{\em Minimal two-band model of the superconducting iron oxypnictides},
\href{http://dx.doi.org/10.1103/PhysRevB.77.220503}{Phys. Rev. B {\bf 77}, 220503 } (2008).

\bibitem{Kee1998}
H.-Y. Kee and C.~M. Varma,
{\em Polarizability and single-particle spectra of two-dimensional $s$- and $d$-wave superconductors},
\href{http://dx.doi.org/10.1103/PhysRevB.58.15035}{Phys. Rev. B {\bf 58}, 15035 } (1998).

\bibitem{Kee1999}
H.-Y. Kee and Y.~B. Kim,
{\em Incommensurate charge and spin fluctuations in $d$-wave superconductors},
\href{http://dx.doi.org/10.1103/PhysRevB.59.4470}{Phys. Rev. B {\bf 59}, 4470 } (1999).

\bibitem{Voo2000}
K.-K. Voo, W.~C. Wu, J.-X. Li, and T.~K. Lee,
{\em Incommensurate charge fluctuations in a $d$-wave superconductor},
\href{http://dx.doi.org/10.1103/PhysRevB.61.9095}{Phys. Rev. B {\bf 61}, 9095 } (2000).

\bibitem{Qureshi2012}
N. Qureshi, P. Steffens, Y. Drees, A.~C. Komarek, D. Lamago, Y. Sidis, L. Harnagea, H.-J. Grafe, S. Wurmehl, B. B\"uchner, and M. Braden,
{\em Inelastic neutron-scattering measurements of incommensurate magnetic excitations on superconducting LiFeAs single crystals},
\href{http://dx.doi.org/10.1103/PhysRevLett.108.117001}{Phys. Rev. Lett. {\bf 108}, 117001 } (2012).

\bibitem{Hess2013}
C. Hess, S. Sykora, T. H\"anke, R. Schlegel, D. Baumann, V.~B. Zabolotnyy, L. Harnagea, S. Wurmehl, J. van~den Brink, and B. B\"uchner,
{\em Interband quasiparticle scattering in superconducting LiFeAs reconciles photoemission and tunneling measurements},
\href{http://dx.doi.org/10.1103/PhysRevLett.110.017006}{Phys. Rev. Lett. {\bf 110}, 017006 } (2013).

\end{thebibliography}
\end{document}